\documentclass[aos,MSNbibl,dvips]{arximspdf}
\usepackage{mathbh}

\doi{10.1214/13-AOS1185} 
\volume{42}
\issue{1}
\pubyear{2014}
\firstpage{171}
\lastpage{189}

\makeatletter
\newcommand{\rrVert}{\Vert}
\newcommand{\llVert}{\Vert}
\newtheorem{theorem}{Theorem}

\newproclaim{definition}[theorem]{Definition}
\newtheorem{lemma}[theorem]{Lemma}
\newcommand{\expect}{\mathrm{E}}
\let\widebar\overline
\makeatother

\begin{document}
\begin{frontmatter}

\title{Second-order asymptotics for quantum hypothesis~testing}
\runtitle{Second-order asymptotics for QHT}

\begin{aug}
\author{\fnms{Ke} \snm{Li}\corref{}\ead[label=e1]{carl.ke.lee@gmail.com}\thanksref{t1}}
\runauthor{K. Li}
\affiliation{IBM TJ Watson Research Center,
Massachusetts Institute of~Technology and
National University of Singapore}
\address{IBM T.~J. Watson Research Center\\
1101 Kitchawan Road \\
Yorktown Heights, New York 10598\\
USA\\
\printead{e1}} 
\end{aug}
\thankstext{t1}{Supported by NSF Grant CCF-1111382. The Centre for
Quantum Technologies is funded by the Singapore Ministry of Education
and the
National Research Foundation as part of the Research Centres of
Excellence program.}

\received{\smonth{1} \syear{2013}}
\revised{\smonth{7} \syear{2013}}

%
\begin{abstract}
In the asymptotic theory of quantum hypothesis testing, the minimal
error probability of
the first kind jumps sharply from zero to one when the error exponent
of the second kind
passes by the point of the relative entropy of the two states in an
increasing way. This
is well known as the direct part and strong converse of quantum Stein's lemma.

Here we look into the behavior of this sudden change and have make it
clear how the error
of first kind grows smoothly according to a lower order of the error
exponent of the second
kind, and hence we obtain the second-order asymptotics for quantum
hypothesis testing.
This actually implies quantum Stein's lemma as a special case.
Meanwhile, our analysis also
yields tight bounds for the case of finite sample size. These results
have potential
applications in quantum information theory.

Our method is elementary, based on basic linear algebra and probability
theory. It deals
with the achievability part and the optimality part in a unified fashion.
\end{abstract}

%
\pdfkeywords{62P35, 62G10, Quantum hypothesis testing,
quantum Stein's lemma, second-order asymptotics, finite sample size}
\end{frontmatter}

\section{Introduction} \label{sectionintroduction}
We are interested in the asymptotic theory of hypothesis testing with
two hypotheses.
Suppose there are many identical physical systems, each independently
being in some
random states, subject to the same statistical description. Here the statistical
description is probability distribution in classical world and quantum
state which is
positive semi-definite matrix with trace 1 in quantum mechanics.
However, the statistical
description is not fixed: it has two possibilities, say, either $\rho$
(the null hypothesis)
or $\sigma$ (the alternative hypothesis). Thus the task is to identify
which statistical
description is the true one, based on the instances of the physical systems.

It is the central problem in asymptotic hypothesis testing to
characterize the behavior
of errors. An intuitive understanding is that the probabilities of
mistaking one
hypothesis for the other can be made arbitrarily small when the sample
size $n$ is big
enough, except for the trivial case that $\rho$ and $\sigma$ are the
same. However,
assuming exponential decay, we want to optimize the rate exponent with
which the error
of concern, under certain reasonable preconditions, converges to zero.
In the classical setting,
this problem has been well understood, featured with a list of famous
results~\cite{Blahut74,Chernoff52,CT91,CsiszarLongo71,HanKobayashi89,Hoeffding65},
including the celebrated Stein's lemma, Chernoff distance and Hoeffding bound.
These results are all obtainable using the likelihood ratio tests.

In contrast to its classical counterpart, the problem of quantum
hypothesis testing becomes
very difficult due to the noncommutativity of the two quantum states
$\rho$ and $\sigma$,
and the more complicated mechanics for observing the underlying
physical systems, that is,
quantum measurement. Although the quantum generalization of the
likelihood ratio test was
obtained in the 1970s~\cite{Helstrom76,Holevo78}, its structure is
not clear in the
aymptotic limit. Yet, substantial achievements have been made since.

In 1991, Hiai and Petz established the quantum Stein's lemma, providing
rigorous operational
interpretation for the quantum relative entropy, or quantum
Kullback--Leibler divergence~\cite{HiaiPetz91}. Then its optimality part was strengthened by Ogawa and
Nagaoka, with a strong
converse theorem~\cite{OgawaNagaoka00}. More recently, quantum
Chernoff distance, the optimal
rate exponent under which the average error tends to $0$ in the setting
of symmetric hypothesis
testing, has been identified in two seminal papers. The achievability
part was due to Audenaert
et al.~\cite{ACMMABV}, and the optimality part was by Nussbaum and
Szko{\l}a~\cite{NussbaumSzkola07}. The methods invented in these two papers were
subsequently used to derive
the quantum Hoeffding bound~\cite{ANSV08,Hayashi06,Nagaoka06}.

The quantum Stein's lemma characterizes the optimal error exponent in
asymmetric hypothesis testing.
Besides the breakthroughs mentioned above, some other important
progresses in this regime can be
found in~\cite{ANSV08,BDKSSS05,BDKSSS08,BS04,BrandaoPlenio09,Hayashi-book,NagaokaHayashi07}.
To state this result, we define two types of errors. Type~I
error (or~the error of the first kind) is the probability that we
incorrectly accept the
alternative hypothesis $\sigma^{\otimes n}$ while it is actually the
null hypothesis
$\rho^{\otimes n}$, and type~II
error (or the error of
the second kind) is the probability of the opposite situation. In an
asymmetric setting,
we want to minimize the type~II
error while only simply
requiring that the type~I error
converges to $0$. Let
$\operatorname{supp}(X)$ be the support of the operator $X$. The
quantum Stein's lemma
states that the maximal
exponent of type~II error is the
quantum relative
entropy~\cite{HiaiPetz91}, given by
\[
D(\rho\|\sigma) = \cases{ \operatorname{Tr} \bigl(\rho(\log\rho -\log\sigma)
\bigr), &\quad if $\operatorname{supp}(\rho)\subseteq\operatorname {supp}(
\sigma)$,
\vspace*{3pt}\cr
+\infty, &\quad otherwise.}
\]
It also asserts that if the type~II
error goes to $0$ with an exponent larger than $D(\rho\|\sigma)$,
then the type~I error inevitably converges to
$1$~\cite{OgawaNagaoka00}.

However, the drawback of the quantum Stein's lemma is that it
characterizes the asymptotic
behavior of errors in a relatively coarse-grained way. To be precise,
it considers only
the linear term of the type~II
error exponent, which
is of the order $n$ (we call it the first order). As a result, the
optimal type~I error jumps sharply from~$0$ to~$1$ when the rate
exponent of type~II
error---quantified by its first
order---passes by the relative entropy $D(\rho\|\sigma)$ from the
smaller side to the larger
side.

In this paper, we prove the second-order asymptotic theorem, and thus
fundamentally refine
the quantum Stein's lemma. Specifically, we track the exponent of the type~II error in depth, to the order
$\sqrt{n}$ (we call
it the second order), and clarify how the type~I error
varies smoothly as a function of this second-order exponent. A
variance-like quantity,
defined as
%
\begin{equation}
\label{eqrelative-variance} V(\rho\|\sigma):=\operatorname{Tr}\rho (\log\rho-\log\sigma
)^2- \bigl(D(\rho\|\sigma) \bigr)^2,
\end{equation}
will play an important role, and we name it the \textit{quantum relative
variance} of $\rho$ and~$\sigma$. Write the second-order rate exponent of the type~II
error as $E_2$. Then our result shows that, asymptotically, the minimal type~I error is given by
$\Phi(E_2/\sqrt{V(\rho\|\sigma)} )$, which grows
smoothly from $0$ to $1$ when
$E_2$ increases from $-\infty$ to $+\infty$. Here $\Phi$ is the
cumulative distribution
function of standard normal distribution, and its appearance in our
result comes from the
use of the central limit theorem in the proof.

We also obtain very tight bounds for the case of finite sample size
$n$. Supposing that the
type~I error is no larger than
a constant $\varepsilon$,
we minimize the type~II error
and consider its negative
logarithm. Then we derive upper and lower bounds for this quantity,
based on the method for
proving our second-order asymptotic theorem. This enables us to
establish that this quantity
can be written as
\[
nD(\rho\|\sigma)+\sqrt{n}\sqrt{V(\rho\|\sigma)}\Phi^{-1}(
\varepsilon)+O(\log n).
\]
The first two terms coincide with the results of the quantum Stein's
lemma and our second-order asymptotic theorem, respectively. Furthermore, the next leading
term (this is the term
of the third order), included in $O(\log n)$ of the above formula, lies
between a constant
and $2\log n$.

Our results have potential applications in quantum information theory.
There is a deep
connection between hypothesis testing and other topics in information
theory (e.g.,
channel capacity), both in the classical regime~\cite{Han-book,VerduHan94} and in the
quantum regime~\cite{HayashiNagaoka03}. Recently, this connection has
been generalized
to the one-shot scenario as well~\cite{MosonyiDatta08,WangRenner10}.
Indeed, such a
connection is very helpful in the derivation of the second-order coding
rate and finite
blocklength analysis in classical channel coding~\cite{Hayashi09,PPV10}. Our results make
it possible to investigate the second-order and finite blocklength
analysis for classical
information transmission over quantum channels.

We point out that the results presented here are independently and
concurrently obtained
by Tomamichel and Hayashi~\cite{TomamichelHayashi12}, using a
different method.
In~\cite{TomamichelHayashi12}, such analysis is conducted in the
context of one-shot
entropies and has been applied to the tasks of data compression with
quantum side
information and randomness extraction against quantum side information.
The bounds for
finite sample size in these two works are slightly different; see
Section~\ref{sectionfinite-sample-size} for details.

The remainder of this paper is organized as follows. In Section~\ref{sectionsecond-order-asymptotics},
we present our main result of second-order asymptotics. Then we prove
it in
Section~\ref{sectionproof}. In Section~\ref{sectionfinite-sample-size}, we treat
the case of finite sample size. In Section~\ref{sectionremarks}, we
note a few remarks.
Finally, we give the proofs to technical lemmas in the \hyperref[appA]{Appendix}.

\section{Second-order asymptotics}\label{sectionsecond-order-asymptotics}
Every quantum system is associated with a~complex Hilbert space. The
state of the quantum
system is described by a density matrix $\varpi$, which is a
nonnegative definite matrix
in the Hilbert space and satisfies the normalization condition
$\operatorname{Tr}\varpi=1$. To detect
the quantum system, we have to do quantum measurement, which, in the
most general form,
is formulated as positive operator-valued measurement (POVM) $\mathcal
{M}=\{M_i\}_i$, with
$0\leq M_i \leq\mathbh{1}$ and $\sum_i M_i=\mathbh{1}$. Then the
measurement outcome $i$ is obtained with
probability $\operatorname{Tr}(\varpi M_i)$.

We consider a large number $n$ of identical quantum systems, each of
which has finite level
and is associated with the Hilbert space $\mathcal{H}$ of finite
dimension $|\mathcal{H}|$. Given that the
quantum systems are either of the state $\rho^{\otimes n}$ (the null
hypothesis) or of the
state $\sigma^{\otimes n}$ (the alternative hypothesis), we want to
identify which state
the systems belong to. Without loss of generality, this can be done by
applying a two-outcome
POVM $(A_n, \mathbh{1}-A_n)$, with $0\leq A_n\leq\mathbh{1}$, on the
joint Hilbert space ${\mathcal{H}}^{\otimes n}$
of the quantum systems. If we obtain the outcome associated to $A_n$,
then we conclude that
the state is $\rho^{\otimes n}$. Similarly, the outcome associated to
$(\mathbh{1}-A_n)$ corresponds
to the state $\sigma^{\otimes n}$. The error probabilities of the
first kind and the second
kind are, respectively, given by $\alpha_n(A_n)=\operatorname
{Tr}(\rho^{\otimes n}(\mathbh{1}-A_n))$ and
$\beta_n(A_n)=\operatorname{Tr}(\sigma^{\otimes n}A_n)$.

The quantum Stein's lemma shows that the relative entropy $D(\rho\|
\sigma)$ is a critical
jump point in the asymptotics of asymmetric hypothesis testing.
Explicitly, it is stated
in two parts as follows:
\begin{itemize}
\item Direct part~\cite{HiaiPetz91}: for arbitrary $R\leq D(\rho\|
\sigma)$, there exist
tests $\{(A_n, \mathbh{1}-A_n)\}_n$ satisfying
\[
\liminf_{n\rightarrow\infty}\frac{-1}{n}\log\beta_n(A_n)
\geq R\quad\mbox{and}\quad\lim_{n\rightarrow\infty}\alpha_n(A_n)
= 0.
\]
\item Strong converse~\cite{OgawaNagaoka00}: if a sequence of tests $\{
(A_n, \mathbh{1}-A_n)\}_n$
is such that
\[
\liminf_{n\rightarrow\infty} \frac{-1}{n}\log\beta_n(A_n)
> D(\rho\|\sigma),
\]
then $\lim_{n\rightarrow\infty}\alpha_n(A_n)=1$.
\end{itemize}

Instead of the rate exponent $\frac{-1}{n}\log\beta_n(A_n)$
considered in the quantum Stein's lemma, we are concerned with a smaller order of the type~II
error exponent, that is, $\frac{1}{\sqrt{n}} (-\log\beta
_n(A_n) - nD(\rho\|\sigma) )$.
Then we think about the optimal tradeoff between the asymptotic limit
of this quantity
and the type~I error $\alpha
_n(A_n)$. In an equivalent
way, we define the error-dependency functions as follows and present
our result subsequently.

%
\begin{definition}
\label{defError-Function}
Let $E_1, E_2\in\mathbb{R}$, and $f(n)$ be a fixed function of some
order other than $n$ or
$\sqrt{n}$, which is to be specified when necessary. We define a
sequence of functions
$\{\alpha_n(E_1,E_2|f)\dvtx  n\in\mathbb{N}\}$, which reflects the
dependency of the minimal
error probability of the first kind on the error exponent of the second
kind, up to the
order $n$ and $\sqrt{n}$, as
\[
\label{eqError-Function} \alpha_n(E_1, E_2|f):=\min
_{A_n} \bigl\{ \alpha_n(A_n) |
\beta_n(A_n)\leq\exp \bigl(- \bigl(E_1n+E_2
\sqrt{n}+f(n) \bigr) \bigr) \bigr\}.
\]
\end{definition}

If $\operatorname{supp}(\rho)\nsubseteq\operatorname{supp}(\sigma
)$, we have $D(\rho\|\sigma
)=+\infty$. Asymptotically,
the optimal error probability of the first kind is always $0$, while
the error exponent of
the second kind can be arbitrarily large. In such a case, the second-order asymptotics makes
no sense. So, in this paper, we suppose $\operatorname{supp}(\rho
)\subseteq\operatorname{supp}
(\sigma)$, and without
loss of generality, we further suppose $\sigma$ is of full rank.

Our main result is the following theorem.

%
\begin{theorem}
\label{thmSecond-Order-Asymptotics}
Let $\{\alpha_n(E_1,E_2|f)\}_n$, the sequence of error-dependency functions,
be as defined in Definition~\ref{defError-Function}, and let $V(\rho
\|\sigma)$, the
quantum relative variance of $\rho$ and $\sigma$, be as defined by
equation~(\ref{eqrelative-variance}).
We have
%
\begin{eqnarray}
\label{eqSecond-Order-Asymptotics}
\lim_{n\rightarrow\infty
}\alpha_n(E_1,
E_2|f)
&=& \cases{ 0, &\quad if $E_1<D(\rho\|\sigma), f\in o(n)$,
\vspace*{3pt}\cr
\Phi
\biggl(\dfrac{E_2}{\sqrt{V(\rho\|\sigma)}} \biggr), &\quad if $E_1=D(\rho\|\sigma)$, $f
\in o(\sqrt{n})$,
\vspace*{3pt}\cr
1, &\quad if $E_1>D(\rho\|\sigma)$, $f\in
o(n)$,}\hspace*{-30pt}
\end{eqnarray}
where $\Phi(x)$ is the cumulative distribution function of the
standard normal distribution,
that is, $\Phi(x):=\frac{1}{\sqrt{2\pi}}\int_{-\infty
}^xe^{-t^2/2}\,\mathrm{d}t$.
\end{theorem}

The second case of equation~(\ref{eqSecond-Order-Asymptotics}) is our
second-order asymptotics.
In fact, it implies the first and third cases, which are nothing else
but the direct part
and strong converse of quantum Stein's lemma, respectively. We include
them here such
that one easily gets the full information at first sight. To see this,
we take the first
case, for example. It is obvious from Definition~\ref
{defError-Function} that, for arbitrary
$E_1<D(\rho\|\sigma)$, $E_2\in\mathbb{R}$, $E_2'\in\mathbb{R}$,
$f(n)\in o(n)$ and
$f'(n)\in o(\sqrt{n})$,
%
\begin{equation}
\label{eqonesmalltwo} \lim_{n\rightarrow\infty}\alpha_n(E_1,
E_2|f)\leq\lim_{n\rightarrow\infty}\alpha_n \bigl(D(
\rho\|\sigma), E_2'|f' \bigr).
\end{equation}
Assuming\vspace*{-1pt} the second case of equation~(\ref
{eqSecond-Order-Asymptotics}), the right-hand side of
equation~(\ref{eqonesmalltwo}) equals $\Phi(\frac{E_2'}{\sqrt {V(\rho\|\sigma)}} )$.
Now\vspace*{1pt} letting $E_2'\rightarrow-\infty$, the first case of
equation~(\ref{eqSecond-Order-Asymptotics})
follows immediately since $\alpha_n(E_1, E_2|f)$ is always nonnegative.

We divide Theorem~\ref{thmSecond-Order-Asymptotics} (precisely, its
second case) into the
achievability part and optimality part, and equivalently reformulate it
below. This
reformulation form corresponds to the structure of the proof in the
next section.

\begin{reformulation*}
For quantum hypothesis testing with the null hypothesis $\rho^{\otimes
n}$ and the alternative
hypothesis $\sigma^{\otimes n}$ and the error probabilities of the
first and second kinds
denoted as $\alpha_n(A_n)$ and $\beta_n(A_n)$, respectively, we have:

Achievability: for any $E_2\in\mathbb{R}$ and $f(n)\in o(\sqrt{n})$,
there exists a
sequence of measurements $\{(A_n, \mathbh{1}-A_n)\}_n$, such that
%
\begin{eqnarray}
\label{eqachi-opti-1} \beta_n(A_n)&\leq&\exp \bigl\{- \bigl(n D(
\rho\|\sigma)+E_2\sqrt{n}+f(n) \bigr) \bigr\},
\\
\label{eqachi-opti-2} \limsup_{n\rightarrow\infty}\alpha_n(A_n)
&\leq&\Phi \biggl(\frac
{E_2}{\sqrt{V(\rho\|\sigma)}} \biggr).
\end{eqnarray}
Optimality: if there is a sequence of measurements $\{(A_n, \mathbh
{1}-A_n)\}_n$ such that
%
\begin{equation}
\label{eqachi-opti-3} \beta_n(A_n)\leq\exp \bigl\{- \bigl(nD(
\rho\|\sigma)+E_2\sqrt{n}+f(n) \bigr) \bigr\}
\end{equation}
holds for given $E_2\in\mathbb{R}$ and $f(n)\in o(\sqrt{n})$, then
%
\begin{equation}
\label{eqachi-opti-4} \liminf_{n\rightarrow\infty}\alpha_n(A_n)
\geq\Phi \biggl(\frac
{E_2}{\sqrt{V(\rho\|\sigma)}} \biggr).
\end{equation}
\end{reformulation*}

The equivalence is obvious. By the definition of $\alpha_n(E_1,
E_2|f)$, it is
straightforward to see that the achievability part of the above
reformulation is
equivalent to
%
\begin{equation}
\label{achiva} \limsup_{n\rightarrow\infty}\alpha_n \bigl(D(\rho
\| \sigma), E_2|f \bigr)\leq\Phi \biggl(\frac{E_2} {
\sqrt{V(\rho\|\sigma)}} \biggr)\qquad \forall f(n)\in o(\sqrt{n})
\end{equation}
and the optimality part of this reformulation is equivalent to
%
\begin{equation}
\label{optima} \liminf_{n\rightarrow\infty}\alpha_n \bigl(D(\rho
\| \sigma), E_2|f \bigr)\geq\Phi \biggl(\frac{E_2} {
\sqrt{V(\rho\|\sigma)}} \biggr)\qquad \forall f(n)\in o(\sqrt{n}).
\end{equation}
Equations~(\ref{achiva}) and (\ref{optima}), in turn, are equivalent
to the second case of
equation~(\ref{eqSecond-Order-Asymptotics}).

\section{Proof of main result}\label{sectionproof}
This section is devoted to the proof of our second-order asymptotics
presented in
Section~\ref{sectionsecond-order-asymptotics}. The proof goes along
the line of the
reformulation of Theorem~\ref{thmSecond-Order-Asymptotics}. At first
we make some necessary
preparations, and then we accomplish the proof by showing the
achievability part and the optimality
part sequentially.

\subsection{Preparations}\label{subsectionproof-prep}
Write $\rho=\sum_x\lambda(x)\vert a_x\rangle\langle a_x \vert$
and $\sigma=\sum_y\mu
(y)\vert b_y\rangle\langle b_y \vert$ in their
diagonal form, where $\{|a_x\rangle\}_x$ and $\{|b_y\rangle\}_y$,
each being an orthonormal
basis of the underlying Hilbert space $\mathcal{H}$, are the
eigenvectors of $\rho$ and $\sigma$,
respectively. $\lambda(x)$ and $\mu(y)$ are the corresponding
eigenvalues, which satisfy
$0\leq\lambda(x)\leq1$, $0<\mu(y)\leq1$ and $\sum_x\lambda
(x)=\sum_y\mu(y)=1$.
Recall that we suppose $\sigma$ is of full rank, and thus $\mu(y)\neq0$.
Let $x^n$ denote the sequence $x_1x_2\ldots x_n$ and $y^n$ denote
$y_1y_2\ldots y_n$. For
$n$ copies of the states $\rho$, we can write
%
\begin{equation}
\label{eqform-rho} \rho^{\otimes n}=\sum_{x^n}
\lambda^n \bigl(x^n \bigr) \bigl\vert a^n_{x^n}
\bigr\rangle \bigl\langle a^n_{x^n} \bigr\vert
\end{equation}
with $\lambda^n(x^n)=\prod_{i=1}^n\lambda(x_i)$ and
$|a^n_{x^n}\rangle=|a_{x_1}\rangle\otimes|a_{x_2}\rangle\otimes\cdots
\otimes|a_{x_n}\rangle$. Similarly,
%
\begin{equation}
\label{eqform-sigma} \sigma^{\otimes n}=\sum_{y^n}
\mu^n \bigl(y^n \bigr) \bigl\vert b^n_{y^n}
\bigr\rangle \bigl\langle b^n_{y^n} \bigr\vert
\end{equation}
with $\mu^n(y^n)=\prod_{i=1}^n\mu(y_i)$ and
$|b^n_{y^n}\rangle=|b_{y_1}\rangle\otimes|b_{y_2}\rangle\otimes\cdots
\otimes|b_{y_n}\rangle$. The subscripts
of $x$ and $y$ indicate which systems they belong to. We further write
$|a_x\rangle$'s
as superpositions of the vectors $\{|b_y\rangle\}_y$, namely,
$|a_x\rangle=\sum_y\gamma_{xy}
|b_y\rangle$, with $\gamma_{xy}=\langle b_y | a_x \rangle\in\mathbb
{C}$ and $\sum_x|\gamma_{xy}|^2=
\sum_y|\gamma_{xy}|^2=1$. In such a way, we have
%
\begin{equation}
\label{eqtransform} \bigl|a^n_{x^n}\bigr\rangle=\sum
_{y^n}\gamma_{x^ny^n}^n\bigl|b^n_{y^n}
\bigr\rangle\qquad\mbox{with } \gamma_{x^ny^n}^n=\prod
_{i=1}^n\gamma_{x_iy_i}.
\end{equation}

Define\vspace*{-1pt} a pair of random variables $(X,Y)$, with alphabet $\{(x,y)\}
_{x,y=1}^{|\mathcal{H}|}$ and
joint distribution $P_{X,Y}(x,y)=\lambda(x)|\gamma_{xy}|^2$.
Operationally, this is the probability of obtaining $(x,y)$ when we
measure the quantum
state $\rho$, sequentially in the bases $\{|a_x\rangle\}_x$ and $\{
|b_y\rangle\}_y$. Let
$(X^n, Y^n):=(X_1,Y_1)(X_2,Y_2)\cdots(X_n,Y_n)$ be a sequence of
independent and identically
distributed random variable pairs, and each $(X_i, Y_i)$ has the same
distribution as
$(X,Y)$. Then
%
\begin{equation}
\label{eqprobability} P_{X^n,Y^n} \bigl(x^n,y^n \bigr)=
\prod_{i=1}^n\lambda(x_i)|
\gamma_{x_iy_i}|^2 =\lambda^n \bigl(x^n
\bigr)\bigl|\gamma_{x^ny^n}^n\bigr|^2.
\end{equation}

As functions of $X$ and $Y$, $\lambda(X)$ and $\mu(Y)$ are also
random variables, and so
are $\lambda^n(X^n)$ and $\mu^n(Y^n)$. Using the idea of Nussbaum and
Szko{\l}a~\cite{NussbaumSzkola07}, we are able to express the quantum
relative entropy
and quantum relative variance as statistical quantities of classical
random variables,
as follows.

%
\begin{lemma}
\label{lemmaCQ-Duality}
We have
%
\begin{eqnarray}
D(\rho\|\sigma)&=&\expect_{(X,Y)}\log\frac{\lambda(X)}{\mu(Y)}, \label{eqR-E-E}
\\
V(\rho\|\sigma)&=&\operatorname{Var}_{(X,Y)}\log\frac{\lambda
(X)}{\mu(Y)}.
\label{eqR-V-V}
\end{eqnarray}
\end{lemma}

Note again that we are only interested in the case that $\sigma$ has
full rank, so
$\mu(Y)>0$. During the computation of the right-hand sides of
equations~(\ref{eqR-E-E}) and~(\ref{eqR-V-V}), if $\lambda(x)=0$, we let
$\lambda(x)\log\lambda(x):=\lim_{z\rightarrow0}z\log z=0$, and\break
$\lambda(x)\log^2\lambda(x):=\lim_{z\rightarrow0}z\log^2 z=0$.

We also present below another technical lemma, which will be used in
Section~\ref{subsectionproof-opti} in the proof of the optimality part.
%
\begin{lemma}
\label{lemmavec-proj-near}
Let $|\phi\rangle$ and $|\varphi\rangle$ be normalized vectors in
some Hilbert space.
Let $\pi$ be a projector and $\|\cdot\|$ the 2-norm, that is, $ \|
|\psi\rangle\|:
=\sqrt{\langle\psi| \psi\rangle}$. If $ \||\phi\rangle-\pi
|\phi\rangle\|\leq\varepsilon$,
then
%
\begin{equation}
\label{eqlemma-vpn-1} \bigl\| \bigl(\vert\phi\rangle \langle\phi \vert \bigr)| \varphi
\rangle\bigr\|^2- \bigl\| \bigl(\pi\vert\phi\rangle \langle \phi\vert\pi \bigr)|
\varphi \rangle\bigr\|^2\leq2\sqrt{2} \varepsilon.
\end{equation}
\end{lemma}

The proofs of Lemmas~\ref{lemmaCQ-Duality}~and~\ref{lemmavec-proj-near} are
given in the \hyperref[appA]{Appendix}.

\subsection{Proof of the achievability part}
\label{subsectionproof-achi}
For any fixed $E_2\in\mathbb{R}$ and $f(n)\in o(\sqrt{n})$, let
\[
L_n:=\exp \bigl\{nD(\rho\|\sigma)+E_2\sqrt{n}+f(n) \bigr\}.
\]
Associated with every $x^n$, we define a projector $Q_{x^n}^n$ as
\[
Q_{x^n}^n:=\sum_{y^n\dvtx \lambda^n(x^n)/\mu^n(y^n)\geq L_n} \bigl\vert
b^n_{y^n}\bigr\rangle \bigl\langle b^n_{y^n}\bigr\vert.
\]
Write $|\xi^n_{x^n}\rangle:=Q_{x^n}^n|a^n_{x^n}\rangle$. Referring
to equation~(\ref{eqtransform}),
we have
%
\begin{equation}
\label{eqproof-a-1} \bigl|\xi^n_{x^n}\bigr\rangle=\sum
_{y^n\dvtx \lambda^n(x^n)/\mu^n(y^n)\geq
L_n}\gamma^n_{x^ny^n}\bigl|b^n_{y^n}\bigr\rangle.
\end{equation}
Let $A_n$ be the projector onto the space $S_n$ that is spanned by $\{
|\xi^n_{x^n}\rangle\}_{x^n}$.
We claim that the sequence of measurements $\{(A_n, \mathbh{1}-A_n)\}
_n$ is what we needed: it
satisfies equations~(\ref{eqachi-opti-1}) and (\ref{eqachi-opti-2}).

Arrange all the values of $x^n$ in such a way that the eigenvalues of
$\rho^{\otimes n}$,
$\lambda^n(x^n)$'s are in an increasing order. This gives an ordering
to the vectors
$\{|\xi^n_{x^n}\rangle\}_{x^n}$ as well. Let $g\dvtx \{i\}
_{i=1}^{|\mathcal{H}|^n}\mapsto\{x^n\}$ be the\vadjust{\goodbreak}
bijection mapping the position of $x^n$ to $x^n$ itself, that is, $x^n$
is at the
$g^{-1}(x^n)$th position in the above ordering. Then we have
%
\begin{equation}
\label{eqproof-a-2} \lambda^n \bigl(g(1) \bigr)\leq\lambda^n
\bigl(g(2) \bigr)\leq\cdots\leq\lambda^n \bigl(g \bigl(|
\mathcal{H}|^n \bigr) \bigr).
\end{equation}
Applying a modified Gram--Schmidt orthonormalization process to the
sequence of vectors
\[
\bigl|\xi^n_{g(1)}\bigr\rangle, \bigl|\xi^n_{g(2)}
\bigr\rangle, \bigl|\xi^n_{g(3)}\bigr\rangle, \ldots,\bigl|\xi^n_{g(|\mathcal{H}|^n)}
\bigr\rangle,
\]
we obtain a new sequence of vectors
%
\begin{equation}
\label{eqproof-a-3} \bigl|\hat{\xi}^n_{g(1)}\bigr\rangle, \bigl|\hat{
\xi}^n_{g(2)}\bigr\rangle, \bigl|\hat{\xi}^n_{g(3)}
\bigr\rangle, \ldots,\bigl|\hat{\xi}^n_{g(|\mathcal{H}|^n)}\bigl\rangle.
\end{equation}
The\vspace*{-1pt} modification is that if $|\xi^n_{g(i)}\rangle\in\operatorname
{Span} (\{|\xi^n_{g(j)}\rangle\}
^{i-1}_{j=1} )$ (this includes the case that $|\xi^n_{g(i)}\rangle
=0$), we let
$|\hat{\xi}^n_{g(i)}\rangle=0$. As a result, the set of vectors $\{
|\hat{\xi}^n_{x^n}\rangle\}_{x^n}$
consists of an orthonormal basis of the space $S_n$, plus some zero
vectors. Thus
%
\begin{equation}
\label{eqproof-a-4} A_n=\sum_{x^n} \bigl\vert\hat{
\xi}^n_{x^n}\bigr\rangle \bigl\langle\hat{\xi }^n_{x^n}
\bigr\vert.
\end{equation}
The vectors $\{|\hat{\xi}^n_{x^n}\rangle\}_{x^n}$ have another
property as follows. From the
Gram--Schmidt process, we know that
%
\begin{equation}
\label{eqproof-a-5} \bigl|\hat{\xi}^n_{g(i)}\bigr\rangle=\sum
^i_{j=1}s^n_{ij}\bigl|\xi^n_{g(j)}\bigr\rangle
\end{equation}
for all $1\leq i\leq|\mathcal{H}|^n$, with the coefficients
$s^n_{ij}\in\mathbb{C}$. Further, from
equations~(\ref{eqproof-a-1}), (\ref{eqproof-a-2}), (\ref{eqproof-a-5}), and paying attention to
the definition of $g$, we conclude that
%
\begin{equation}
\label{eqproof-a-6} \bigl|\hat{\xi}^n_{x^n}\bigr\rangle=\sum
_{y^n\dvtx \lambda^n(x^n)/\mu^n(y^n)\geq
L_n}t^n_{x^ny^n}\bigl|b^n_{y^n}\bigr\rangle,
\end{equation}
where $t^n_{x^ny^n}\in\mathbb{C}$ and
%
\begin{equation}
\label{eqproof-a-7} \sum_{y^n\dvtx \lambda^n(x^n)/\mu^n(y^n)\geq
L_n}\bigl|t^n_{x^ny^n}\bigr|^2=1.
\end{equation}
Equations~(\ref{eqform-sigma}), (\ref{eqproof-a-6}), (\ref{eqproof-a-7}) lead to
%
\begin{equation}
\label{eqproof-a-8} \operatorname{Tr} \bigl(\sigma^{\otimes
n} \bigl\vert\hat{\xi
}^n_{x^n}\bigr\rangle \bigl\langle\hat{\xi}^n_{x^n}
\bigr\vert \bigr) = \sum_{y^n\dvtx \lambda^n(x^n)/\mu^n(y^n)\geq
L_n}\bigl|t^n_{x^ny^n}\bigr|^2
\mu^n \bigl(y^n \bigr) \leq\frac{\lambda^n(x^n)}{L_n}.
\end{equation}
So, making use of equations~(\ref{eqproof-a-4}) and (\ref
{eqproof-a-8}), we arrive at
%
\begin{equation}
\label{eqproof-a-9} \beta_n(A_n) = \operatorname{Tr}
\sigma^{\otimes n}A_n \leq\frac{1}{L_n}=\exp \bigl\{- \bigl(nD(
\rho\|\sigma)+E_2\sqrt{n}+f(n) \bigr) \bigr\},
\end{equation}
which is exactly equation~(\ref{eqachi-opti-1}).\vadjust{\goodbreak}

On the other hand, equation~(\ref{eqachi-opti-2}) is confirmed as
follows. Let
\[
\bigl|\bar{\xi}^n_{x^n}\bigr\rangle:= \cases{ 0, &\quad if $\bigl|\xi^n_{x^n}\bigr\rangle=0$,
\vspace*{3pt}\cr
\dfrac{|\xi^n_{x^n}\rangle}{\sqrt{\langle\xi^n_{x^n} | \xi
^n_{x^n} \rangle}}, &\quad if $\bigl|\xi^n_{x^n}\bigr\rangle\neq0$.}
\]
Obviously, $|\bar{\xi}^n_{x^n}\rangle\in S_n$. So
%
\begin{equation}
\label{eqproof-a-10} \bigl\vert\bar{\xi}^n_{x^n}\bigr\rangle \bigl\langle\bar{\xi}^n_{x^n} \bigr\vert\leq A_n.
\end{equation}
Then we have
\begin{eqnarray*}
\alpha_n(A_n)&=&1-\operatorname{Tr} \bigl(
\rho^{\otimes n}A_n \bigr)
\\
&\leq& 1-\sum_{x^n}\lambda^n
\bigl(x^n \bigr)\operatorname{Tr} \bigl( \bigl( \bigl\vert
a^n_{x^n}\bigr\rangle \bigl\langle a^n_{x^n}
\bigr\vert \bigr) \bigl( \bigl\vert\bar{\xi}^n_{x^n} \bigr\rangle
\bigl\langle\bar{\xi}^n_{x^n} \bigr\vert \bigr) \bigr)
\\
&=&1-\sum_{x^n}\lambda^n
\bigl(x^n \bigr) \bigl\langle\xi^n_{x^n} |
\xi^n_{x^n} \bigr\rangle
\\
&=&\operatorname{Pr} \biggl\{\frac{\lambda^n(X^n)}{\mu^n(Y^n)} < L_n \biggr\},
\end{eqnarray*}
where the second line is by equations~(\ref{eqform-rho}) and (\ref
{eqproof-a-10}), the
third line can be seen from the definitions of $|\xi^n_{x^n}\rangle$
and $|\bar{\xi}^n_{x^n}\rangle$ and
the fourth line follows from equations (\ref{eqproof-a-1})~and~(\ref{eqprobability}).
Recalling that $\lambda^n(X^n)=\prod^n_{i=1}\lambda(X_i)$ and $\mu
^n(Y^n)=\prod^n_{i=1}\mu(Y_i)$,
and by taking logarithms at both sides of $\frac{\lambda^n(X^n)}{\mu
^n(Y^n)} < L_n $,
we further obtain
%
\begin{equation}
\label{eqproof-a-11} \alpha_n(A_n)\leq\operatorname{Pr} \Biggl\{
\sqrt{n} \Biggl(\frac{1}{n}\sum^n_{i=1}
\log\frac{\lambda(X_i)}{\mu(Y_i)}-D(\rho\|\sigma) \Biggr)<E_2+\frac
{f(n)}{\sqrt{n}}
\Biggr\}.
\end{equation}
Since $f(n)\in o(\sqrt{n})$, due to the central limit theorem and also
by Lemma~\ref{lemmaCQ-Duality},
the limit of right-hand side of equation~(\ref{eqproof-a-11}) equals
\[
\Phi \biggl(\frac{E_2}{\sqrt{V(\rho\|\sigma)}} \biggr);
\]
thus equation~(\ref{eqachi-opti-2}) follows, and we are done.

\subsection{Proof of the optimality part}
\label{subsectionproof-opti}
Suppose that the sequence of measurements $\{(A_n, \mathbh{1}-A_n)\}
_n$ satisfies
equation~(\ref{eqachi-opti-3}). We will prove equation~(\ref
{eqachi-opti-4}). Let
%
\begin{equation}
\label{eqproof-o-a} L_n:=\exp \bigl\{ \bigl(nD(\rho\|
\sigma)+E_2\sqrt{n}+f(n) \bigr)-f'(n) \bigr\}
\end{equation}
with some fixed
%
\begin{equation}
\label{eqproof-o-b} f'(n)\in o(\sqrt{n})\cap\omega(1).
\end{equation}
Here $\omega(1)$ is the family of functions that are defined on
$\mathbb{N}$ and diverge
to $+\infty$. Associated with every $x^n$, we define the projector
$Q_{x^n}^n$ as
%
\begin{equation}
\label{eqopti-projector} Q_{x^n}^n:=\sum
_{y^n\dvtx \lambda^n(x^n)/\mu^n(y^n)\geq L_n} \bigl\vert b^n_{y^n}\bigr\rangle \bigl\langle
b^n_{y^n} \bigr\vert.
\end{equation}
Inserting equation~(\ref{eqform-rho}) into the definition of $\alpha
_n(A_n)$, namely,
$\alpha_n(A_n):= \operatorname{Tr}(\rho^{\otimes n}(\mathbh
{1}-A_n))$, and after a few calculations, we write
%
\begin{equation}
\label{eqproof-o-c} \alpha_n(A_n)=1-C_n-D_n,
\end{equation}
where $C_n$ and $D_n$ are
%
\begin{eqnarray}
C_n&:=&\sum
_{x^n}\lambda^n
\bigl(x^n \bigr)\operatorname{Tr} \bigl(Q^n_{x^n}
\sqrt{A_n} \bigl\vert a^n_{x^n}\bigr\rangle \bigl\langle
a^n_{x^n} \bigr\vert\sqrt{A_n}Q^n_{x^n}
\bigr), \label{eqproof-o-d}
\\
D_n&:=&\sum
_{x^n}\lambda^n
\bigl(x^n \bigr)\operatorname{Tr} \bigl( \bigl(\mathbh{1}-Q^n_{x^n}
\bigr)\sqrt{A_n} \bigl\vert a^n_{x^n} \bigr\rangle
\bigl\langle a^n_{x^n} \bigr\vert\sqrt{A_n} \bigl(
\mathbh{1}-Q^n_{x^n} \bigr) \bigr). \label{eqproof-o-e}
\end{eqnarray}

The basic difficulty in bounding $C_n$ and $D_n$ is that the POVM
element $A_n$ is very
general, except for the constraint of equation~(\ref{eqachi-opti-3}).
Nevertheless, we will
be able to show that the $D_n$ term is asymptotically negligible, due
to the constraint
of equation~(\ref{eqachi-opti-3}) and our choice of $L_n$. This in
turn, ensures that the
$C_n$ term can be upper bounded by removing the operator ``$\sqrt {A_n}$'' from its
expression, with only an infinitesimal correction; cf. equation~(\ref{eqproof-o-n}).

Now we show that the $D_n$ term is asymptotically negligible. Because
\[
\sigma^{\otimes n}\geq \bigl(\mathbh{1}-Q^n_{x^n} \bigr)
\sigma^{\otimes
n} \bigl(\mathbh{1}-Q^n_{x^n} \bigr)\geq
\frac{\lambda^n(x^n)}{L_n} \bigl(\mathbh{1}-Q^n_{x^n} \bigr),
\]
where the first inequality is owing to the commutativity of $\sigma
^{\otimes n}$ and the projector
$(\mathbh{1}-Q^n_{x^n})$, and the second one can be seen from the
definition of $Q^n_{x^n}$, we obtain
\begin{eqnarray*}
\beta_n(A_n)&=&\operatorname{Tr} \bigl(
\sigma^{\otimes n}A_n \bigr)
\\
&=&\sum_{x^n}
\operatorname{Tr} \bigl(\sigma^{\otimes n} \bigl(\sqrt{A_n} \bigl\vert
a^n_{x^n}\bigr\rangle \bigl\langle a^n_{x^n}
\bigr\vert\sqrt{A_n} \bigr) \bigr)
\\
&\geq&\sum_{x^n}\operatorname{Tr} \biggl(
\frac{\lambda
^n(x^n)}{L_n} \bigl(\mathbh{1}-Q^n_{x^n} \bigr) \bigl(
\sqrt{A_n} \bigl\vert a^n_{x^n} \bigr\rangle \bigl\langle
a^n_{x^n} \bigr\vert\sqrt{A_n} \bigr) \biggr)
\\
&=&\frac{D_n}{L_n}.
\end{eqnarray*}
This result, together with equations~(\ref{eqachi-opti-3}), (\ref
{eqproof-o-a}) and
(\ref{eqproof-o-b}), tells us that
%
\begin{equation}
\label{eqproof-o-f} D_n \leq L_n\beta_n(A_n)
\leq\exp \bigl\{-f'(n) \bigr\} \rightarrow0.
\end{equation}

The evaluation of the $C_n$ term will be a bit more complicated. For
simplicity, we use the
notation of norm, $ \||\psi\rangle\|=\sqrt{\langle\psi|
\psi\rangle}=\sqrt{\operatorname{Tr}\vert \psi\rangle\langle
\psi\vert}$,
with $|\psi\rangle$ being a vector of some Hilbert space. Thus $C_n$
is rewritten as
%
\begin{equation}
\label{eqproof-o-h} C_n=\sum
_{x^n}\lambda^n
\bigl(x^n \bigr) \bigl\| Q^n_{x^n}\sqrt
{A_n}\bigl|a^n_{x^n}\bigr\rangle\bigr\|^2.
\end{equation}
Our strategy is to divide the terms in the sum of the above expression
into different
classes, each satisfying some special conditions. Then we evaluate them
individually under
these conditions. For such a purpose, we define index sets
\begin{eqnarray*}
\mathcal{O}^n_1&:=& \bigl\{x^n | \bigl\|
\sqrt{A_n}\bigl|a^n_{x^n}\bigr\rangle\bigr\|\geq
\epsilon_1 \bigr\},
\\
\mathcal{O}^n_2&:=& \bigl\{x^n | \bigl\| \bigl(
\mathbh{1}-Q^n_{x^n} \bigr)\sqrt{A_n}\bigl|a^n_{x^n}
\bigr\rangle\bigr\|\leq\epsilon_1\epsilon_2 \bigr\}
\end{eqnarray*}
with sufficiently small $\epsilon_1, \epsilon_2>0$. Denote the full
set of all the
$x^n$'s as $\mathcal{O}^n$, and the complementary sets of $\mathcal
{O}^n_1$ and $\mathcal{O}^n_2$ as
$\overline{\mathcal{O}^n_1}$ and $\overline{\mathcal{O}^n_2}$,
respectively. Since $\mathcal{O}^n$ is the union of
three disjoint subsets
\[
\mathcal{O}^n=\overline{\mathcal{O}^n_1}
\cup \bigl(\mathcal{O}^n_1\cap\overline{
\mathcal{O}^n_2} \bigr)\cup \bigl(\mathcal{O}^n_1
\cap\mathcal{O}^n_2 \bigr),
\]
we deal with equation~(\ref{eqproof-o-h}) under distinct cases that
$x^n$ belongs to these subsets,
respectively, and then sum them up.

The first case is that $x^n\in\overline{\mathcal{O}^n_1}$. Noting
that a projector (more generally,
any contraction whose singular values are no larger than 1) acting on a
vector will not
increase its norm, we have
%
\begin{equation}
\label{eqproof-o-i} \sum_{x^n\in\overline{\mathcal
{O}^n_1}}\lambda^n
\bigl(x^n \bigr) \bigl\| Q^n_{x^n}
\sqrt{A_n}\bigl|a^n_{x^n} \bigr\rangle\bigr\|^2 \leq
\sum_{x^n\in\overline{\mathcal{O}^n_1}} \lambda^n \bigl(x^n
\bigr)\epsilon_1^2 \leq\epsilon_1^2.
\end{equation}

The second case is that $x^n\in\mathcal{O}^n_1\cap\overline
{\mathcal{O}^n_2}$. We upper bound it as
%
\begin{eqnarray}
\label{eqproof-o-j} &&\sum_{x^n\in\mathcal{O}^n_1\cap\overline
{\mathcal
{O}^n_2}}\lambda^n
\bigl(x^n \bigr) \bigl\| Q^n_{x^n}
\sqrt{A_n}\bigl|a^n_{x^n} \bigr\rangle\bigr\|^2\nonumber
\\
&&\qquad \leq\sum_{x^n\in\overline{\mathcal{O}^n_2}}\lambda^n
\bigl(x^n \bigr) \leq\sum_{x^n\in\overline{\mathcal{O}^n_2}}
\lambda^n \bigl(x^n \bigr)\frac
{1}{\epsilon_1^2\epsilon_2^2} \bigl\| \bigl(
\mathbh{1}-Q^n_{x^n} \bigr)\sqrt{A_n}\bigl|a^n_{x^n}
\bigr\rangle\bigr\|^2
\\
&&\qquad \leq\frac{1}{\epsilon_1^2\epsilon_2^2}\sum_{x^n}\lambda
^n \bigl(x^n \bigr) \bigl\| \bigl(\mathbh{1}-Q^n_{x^n}
\bigr)\sqrt{A_n}\bigl|a^n_{x^n}\bigr\rangle
\bigr\|^2 = \frac{D_n}{\epsilon_1^2\epsilon_2^2},\nonumber
\end{eqnarray}
where for the first inequality we use $ \| Q^n_{x^n}\sqrt {A_n}|a^n_{x^n}\rangle\|
\leq\| \sqrt{A_n}|a^n_{x^n}\rangle\| \leq\break  \|
|a^n_{x^n}\rangle\|=1$, the second
inequality is by definition of $\mathcal{O}^n_2$ and the last equality
can be easily seen from
equation~(\ref{eqproof-o-e}) and the definition of norm.

The last case, which will turn out to be the dominant part, is that
$x^n\in\mathcal{O}^n_1\cap\mathcal{O}^n_2$.
In such a case, paying attention to the definition of $\mathcal
{O}^n_1$ and $\mathcal{O}^n_2$, we see that
\[
\biggl\llVert\frac{\sqrt{A_n}|a^n_{x^n}\rangle}{ \|\sqrt {A_n}|a^n_{x^n}\rangle\|}-Q^n_{x^n}
\frac{\sqrt{A_n}|a^n_{x^n}\rangle}{ \|\sqrt{A_n}|a^n_{x^n}\rangle\|
} \biggr\rrVert=\frac{ \|(\mathbh{1}-Q^n_{x^n})\sqrt {A_n}|a^n_{x^n}\rangle
\|}{ \|\sqrt{A_n}|a^n_{x^n}\rangle
\|}\leq\frac{\epsilon_1\epsilon_2}{\epsilon_1}=
\epsilon_2.
\]
Then, directly applying Lemma~\ref{lemmavec-proj-near}, we get
%
\begin{eqnarray}
\label{eqproof-o-k} && \biggl\llVert\frac{\sqrt{A_n}\vert
a^n_{x^n}\rangle\langle
a^n_{x^n} \vert\sqrt{A_n}}{ \|\sqrt{A_n}|a^n_{x^n}\rangle
\|^2}\bigl|a^n_{x^n}
\bigr\rangle \biggr\rrVert^2
\nonumber\\[-8pt]\\[-8pt]
&&\qquad \leq \biggl\llVert \biggl(Q^n_{x^n}\frac{\sqrt{A_n}\vert
a^n_{x^n}\rangle\langle a^n_{x^n} \vert\sqrt{A_n}}{ \|\sqrt{A_n}
|a^n_{x^n}\rangle\|^2}Q^n_{x^n}
\biggr)\bigl|a^n_{x^n}\bigr\rangle \biggr\rrVert^2 + 2
\sqrt{2}\epsilon_2.\nonumber
\end{eqnarray}
Since $0\leq A_n \leq\mathbh{1}$, it holds that $A_n\leq\sqrt {A_n}$. As a result,
%
\begin{equation}
\label{eqproof-o-l}\quad  \bigl\|Q^n_{x^n}\sqrt{A_n}\bigl|a^n_{x^n}
\bigr\rangle\bigr\|^2 \leq\bigl\| \sqrt{A_n}\bigl|a^n_{x^n}
\bigr\rangle\bigr\|^2 \leq \biggl\llVert\frac{\langle a^n_{x^n}|\sqrt {A_n}|a^n_{x^n}\rangle
}{\langle a^n_{x^n}|A_n|a^n_{x^n}\rangle}
\sqrt{A_n}\bigl|a^n_{x^n}\bigr\rangle \biggr\rrVert
^2.
\end{equation}
The last term of equation~(\ref{eqproof-o-l}) and the left-hand side
of equation~(\ref{eqproof-o-k}) are
actually the same. So, combining these two equations together, and
noting that the right-hand side
of equation~(\ref{eqproof-o-k}) is obviously upper bounded by
\[
\bigl\|Q^n_{x^n}\bigl|a^n_{x^n}\bigr\rangle
\bigr\|^2 + 2\sqrt{2}\epsilon_2,
\]
we arrive at
%
\begin{eqnarray}
\label{eqproof-o-m} &&\sum_{x^n\in\mathcal{O}^n_1\cap\mathcal
{O}^n_2}\lambda^n
\bigl(x^n \bigr) \bigl\| Q^n_{x^n}
\sqrt{A_n}\bigl|a^n_{x^n} \bigr\rangle\bigr\|^2\nonumber
\\
&&\qquad \leq\sum_{x^n\in\mathcal{O}^n_1\cap\mathcal{O}^n_2}\lambda^n
\bigl(x^n \bigr) \bigl( \bigl\|Q^n_{x^n}\bigl|a^n_{x^n}
\bigr\rangle\bigr\|^2 + 2\sqrt{2}\epsilon_2 \bigr)
\\
&&\qquad \leq \sum_{x^n}\lambda^n
\bigl(x^n \bigr) \bigl\| Q^n_{x^n}\bigl|a^n_{x^n}
\bigr\rangle\bigr\|^2 + 2\sqrt{2}\epsilon_2.\nonumber
\end{eqnarray}

Now, adding equations~(\ref{eqproof-o-i}), (\ref{eqproof-o-j}) and
(\ref{eqproof-o-m}) together,
we obtain from equation~(\ref{eqproof-o-h}) that
%
\begin{equation}
\label{eqproof-o-n} C_n \leq\sum_{x^n}
\lambda^n \bigl(x^n \bigr) \bigl\|Q^n_{x^n}\bigl|a^n_{x^n}
\bigr\rangle\bigr\|^2 + \frac{D_n}{\epsilon_1^2\epsilon_2^2}+\epsilon_1^2+2
\sqrt{2}\epsilon_2.
\end{equation}

In analogy to the process in the derivation of equation~(\ref
{eqachi-opti-2}) in
Section~\ref{subsectionproof-achi}, making use of equations~(\ref
{eqtransform}), (\ref{eqopti-projector})
and then (\ref{eqprobability}), we can check that the first term of
the right-hand side of
equation~(\ref{eqproof-o-n}) is equal to the probability of the event
$\{\lambda^n(X^n)/\mu^n(Y^n)
\geq L_n\}$, which is equivalent to
\[
\Biggl\{\sqrt{n} \Biggl(\frac{1}{n}\sum^n_{i=1}
\log\frac{\lambda
(X_i)}{\mu(Y_i)}-D(\rho\|\sigma) \Biggr) \geq E_2+
\frac{f(n)-f'(n)}{\sqrt{n}} \Biggr\}.
\]
So inserting equations~(\ref{eqproof-o-f}) and (\ref{eqproof-o-n})
into equation~(\ref{eqproof-o-c}),
we eventually obtain
%
\begin{eqnarray}
\label{eqproof-o-o} \qquad \alpha_n(A_n) &\geq&\operatorname{Pr} \Biggl\{
\sqrt{n} \Biggl(\frac{1}{n}\sum^n_{i=1}
\log\frac{\lambda(X_i)}{\mu(Y_i)}- D(\rho\|\sigma) \Biggr)\leq E_2+
\frac{f(n)-f'(n)}{\sqrt{n}} \Biggr\}
\nonumber\\[-8pt]\\[-8pt]
&&{}- \biggl(\frac{1}{\epsilon_1^2\epsilon_2^2}+1 \biggr)\exp \bigl\{ -f'(n) \bigr\} -
\epsilon_1^2-2\sqrt{2}\epsilon_2.\nonumber
\end{eqnarray}
Recalling that $f(n)\in o(\sqrt{n})$ and $f'(n)\in o(\sqrt{n})\cap
\omega(1)$, and then
making use of the central limit theory and Lemma~\ref{lemmaCQ-Duality}, we see that the
right-hand side of equation~(\ref{eqproof-o-o}) converges to
\[
\Phi \biggl(\frac{E_2}{\sqrt{V(\rho\|\sigma)}} \biggr)-\epsilon_1^2-2
\sqrt{2}\epsilon_2,
\]
when $n\rightarrow\infty$. Thus equation~(\ref{eqachi-opti-4})
follows, since $\epsilon_1$
and $\epsilon_2$ can be arbitrarily small, and we are done.

\section{Finite sample size analysis}
\label{sectionfinite-sample-size}
In Section~\ref{sectionproof}, we proved the second-order
asymptotics. Here we show that
our method is able to provide tight bounds for the case of finite
sample size as well. The
basic idea is to use the Berry--Esseen theorem instead of the central
limit theorem.

The Berry--Esseen theorem quantifies how fast the standardized mean of
a random sample converges
to a normal distribution. Let $X_1, X_2,\ldots, X_n$ be i.i.d. random~variables, with
$\expect(X_i)=\widebar{X}$, $\expect(X_i-\widebar{X})^2=\varrho^2>0$, and
$\expect|X_i-\widebar{X}|^3=
\varsigma^3<+\infty$. Then it asserts
%
\begin{equation}
\label{eqberry-esseen} \Biggl\vert\operatorname{Pr} \Biggl\{\sqrt{n} \Biggl( \frac{1}{n}\sum
_{i=1}^nX_i- \widebar{X} \Biggr)
\leq x \Biggr\} -\Phi \biggl(\frac{x}{\varrho} \biggr) \Biggr\vert\leq
\frac{C\varsigma
^3}{\sqrt{n}\varrho^3},
\end{equation}
where $0.40973\leq C \leq0.4784$ is a constant~\cite{Korolev-Shevtsova2012}.

Consider the minimal type~II
error given that the type~I error is no larger than some
constant, and define
$\beta_n(\varepsilon):=\min_{A_n} \{\beta_n(A_n)
|\alpha_n(A_n)\leq\varepsilon\}$.
Theorem~\ref{thmFinite-Sample-Size} provides this quantity with tight upper
and lower bounds. It has fixed the second-order term in the asymptotic
expansion of
$-\log\beta_n(\varepsilon)$, and also indicates that the third-order
term lies between a
constant and $2\log n$. Note that previously, only the first-order term
($nD(\rho\|\sigma)$)
is known exactly~\cite{HiaiPetz91,OgawaNagaoka00}, and the
second-order term is known to
be of the order $\sqrt{n}$~\cite{AMV12}. Compared to the upper and
lower bounds
obtained in the independent work of Tomamichel and Hayashi~\cite
{TomamichelHayashi12}, those presented here
are tighter in the third-order term and are also relatively cleaner
because those in~\cite{TomamichelHayashi12} depend on more parameters, such as the
number of distinct
eigenvalues and the ratio between the maximum and minimum eigenvalues
of the quantum state.

%
\begin{theorem}
\label{thmFinite-Sample-Size}
Let $C$ be the constant in the Berry--Esseen theorem, and let
$T^3=\expect_{(X,Y)}\vert\log
\frac{\lambda(X)}{\mu(Y)}-D(\rho\|\sigma)\vert^3$; cf.
Section~\ref{subsectionproof-prep}.
Then for $n$ sufficiently large such that $\varepsilon-\frac{1}{\sqrt {n}}\frac{CT^3} {
\sqrt{V(\rho\|\sigma)}^3}\geq0$, we have
%
\begin{eqnarray}
\label{eqfinite-lower-bound}
-\log\beta_n(\varepsilon)&\geq&
nD(\rho\|\sigma)+\sqrt{n}\sqrt{V(\rho\|\sigma)}\Phi^{-1} \biggl(
\varepsilon-\frac{1} {
\sqrt{n}}\frac{CT^3}{\sqrt{V(\rho\|\sigma)}^3} \biggr)
\nonumber\\[-9pt]\\[-9pt]
&=& nD(\rho\|\sigma)+\sqrt{n}\sqrt{V(\rho\|\sigma)}\Phi^{-1}(
\varepsilon)+O(1)\nonumber
\end{eqnarray}
and for $n$ sufficiently large such that
$\varepsilon+\frac{1}{\sqrt{n}} (\frac{CT^3}{\sqrt{V(\rho\|
\sigma)}^3}+2 )\leq1$,
we have
%
\begin{eqnarray}
\label{eqfinite-upper-bound}
-\log\beta_n(\varepsilon)&\leq& nD(\rho\|\sigma)
+\sqrt{n}\sqrt{V(\rho\|\sigma)}\Phi^{-1} \biggl(
\varepsilon+\frac{1} {
\sqrt{n}} \biggl(\frac{CT^3}{\sqrt{V(\rho\|\sigma)}^3}+2 \biggr) \biggr)\hspace*{-25pt}\nonumber
\\
&&{} +\log \bigl(2^9n^2 \bigr)
\\
&=& nD(\rho\|\sigma)+\sqrt{n}\sqrt{V(\rho\|\sigma)}\Phi^{-1}(
\varepsilon)+2\log n+O(1).\nonumber
\end{eqnarray}
\end{theorem}

\begin{pf}
The equalities in equations~(\ref{eqfinite-lower-bound}) and (\ref
{eqfinite-upper-bound}) are
easy to see by expanding $\Phi^{-1}$ at the point $\varepsilon$ using
Lagrange's mean
value theorem. So it suffices to prove the two inequalities.

Applying the Berry--Esseen theorem to the right-hand side of
equation~(\ref{eqproof-a-11}), and then
following from the argument in Section~\ref{subsectionproof-achi}
[cf. equations~(\ref{eqproof-a-9})
and (\ref{eqproof-a-11})], we have for any $E_2\in\mathbb{R}$ and
$f(n)\in o(\sqrt{n})$,
there exists a sequence of measurements $\{(A_n, \mathbh{1}-A_n)\}_n$,
such that
%
\begin{eqnarray}
\alpha_n(A_n)&\leq&\Phi \biggl(\frac{E_2+f(n)/\sqrt{n}}{\sqrt {V(\rho\|\sigma)}}
\biggr)+ \frac{1}{\sqrt{n}}\frac{CT^3}{\sqrt{V(\rho\|\sigma
)}^3}, \label
{eqfinite-1}
\\
\beta_n(A_n)&\leq&\exp \bigl\{- \bigl(n D(\rho\|
\sigma)+E_2\sqrt{n}+f(n) \bigr) \bigr\}. \label{eqfinite-2}
\end{eqnarray}
When $\varepsilon-\frac{1}{\sqrt{n}}\frac{CT^3}{\sqrt{V(\rho\|
\sigma)}^3}\geq0$, letting the
right-hand side of equation~(\ref{eqfinite-1}) be equal to
$\varepsilon$, then eliminating $E_2\sqrt{n}+f(n)$
from equation~(\ref{eqfinite-2}), we get
\begin{eqnarray*}
\alpha_n(A_n)&\leq&\varepsilon,
\\
\beta_n(A_n)&\leq&\exp \biggl\{- \biggl(n D(\rho\|
\sigma)+ \sqrt{n}\sqrt{V(\rho\|\sigma)}\Phi^{-1} \biggl(\varepsilon-
\frac{1}{\sqrt{n}}\frac{CT^3}{\sqrt{V(\rho\|
\sigma)}^3} \biggr) \biggr) \biggr\}.
\end{eqnarray*}
This, by the definition of $\beta_n(\varepsilon)$, leads to the first
inequality in
equation~(\ref{eqfinite-lower-bound}).

On the other hand, applying the Berry--Esseen theorem to the first term
of the right-hand side of
equation~(\ref{eqproof-o-o}), then the argument in Section~\ref{subsectionproof-opti} implies the following
[cf. the precondition and equation~(\ref{eqproof-o-o})]: if there is
a sequence of measurements
$\{(A_n, \mathbh{1}-A_n)\}_n$ such that
\[
\beta_n(A_n)\leq\exp \bigl\{- \bigl(n D(\rho\|
\sigma)+E_2\sqrt{n}+f(n) \bigr) \bigr\},
\]
then
\[
\alpha_n(A_n)\geq\Phi \biggl(\frac{E_2+(f(n)-f'(n))/\sqrt {n}}{\sqrt{V(\rho\|\sigma)}}
\biggr)-F
\]
with $F=\frac{1}{\sqrt{n}}\frac{CT^3}{\sqrt{V(\rho\|\sigma
)}^3}+ (\frac{1} {
\epsilon_1^2\epsilon_2^2}+1 )\exp\{-f'(n)\}+\epsilon
_1^2+2\sqrt{2}\epsilon_2$.
Since $\alpha_n$ and $\beta_n$ are continuous functionals of $A_n$,
this equivalently
states that if
%
\begin{equation}
\label{eqfinite-3} \alpha_n(A_n)\leq\Phi \biggl(
\frac{E_2+(f(n)-f'(n))/\sqrt{n}}{\sqrt{V(\rho\|\sigma)}} \biggr)-F,
\end{equation}
then
%
\begin{equation}
\label{eqfinite-4} \beta_n(A_n)\geq\exp \bigl\{- \bigl(n D(
\rho\|\sigma)+E_2\sqrt{n}+f(n) \bigr) \bigr\}.
\end{equation}
When $\varepsilon+F \leq1$, let $E_2$ and $f(n)$ be such that the
right-hand side of equation~(\ref{eqfinite-3})
equals $\varepsilon$, then we eliminate $E_2\sqrt{n}+f(n)$ from
equation~(\ref{eqfinite-4}) using
this equality. Thus the above statement implies, by the definition of
$\beta_n(\varepsilon)$,
%
\begin{equation}
\label{eqfinite-5} \beta_n(\varepsilon)\geq\exp \bigl\{- \bigl(n D(\rho
\|\sigma)+\sqrt{n}\sqrt{V(\rho\|\sigma)}\Phi^{-1} (\varepsilon+F
)+f'(n) \bigr) \bigr\}.
\end{equation}
At last, to optimize over the parameters, let $\epsilon_1=2^{1/8}\exp\{-\frac{1}{8}f'(n)\}$,
$\epsilon_2=2^{-1/4}\exp\{-\frac{1}{4}f'(n)\}$ and
$f'(n)=\log(2^9n^2)$. Thus
$F \leq\frac{1}{\sqrt{n}} (\frac{CT^3}{\sqrt{V(\rho\|\sigma
)}^3}+2 )$. Inserting
these into equation~(\ref{eqfinite-5}) results in the inequality of
equation~(\ref{eqfinite-upper-bound}),
and we are done.
\end{pf}

\section{Concluding remarks}
\label{sectionremarks}
The relation between our second-order asymptotics and the quantum Stein's lemma is similar in spirit to the relation between the central limit
theorem and the week law of large numbers. Indeed, we have
employed the central limit theorem and the Berry--Esseen theorem, to
derive our results.

We have succeeded in proving the results using elementary linear
algebra and probability theory
in a unified fashion for the achievability part and optimality part.
Specifically, we have explicitly
constructed a sequence of asymptotically optimal tests for our problem,
specifying the bases of
spaces onto which the projective measurements are applied by employing
a modified Gram--Schmidt
orthonormalization process. In~\cite{NussbaumSzkola11}, the
Gram--Schmidt orthonormalization
process has already been used in order to find the asymptotically
optimal tests for testing multiple
hypotheses in the symmetric setting (regarding the Chernoff bound). The
attempt in~\cite{NussbaumSzkola11}
is successful in some special cases, and is successful in general up to
a constant factor $1/3$. We notice that
even for two hypotheses, such an elementary method is not known for
fully proving the achievability of the
Chernoff bound; recall that the original---and hitherto unique---proof in~\cite{ACMMABV} was based on
the nontrivial matrix inequality $\operatorname{Tr}(\rho^s\sigma
^{1-s})\geq\operatorname{Tr}(\rho+\sigma-|\rho-\sigma|)/2$, for
all $0 \leq s \leq1$.

The case that $V(\rho\|\sigma)=0$ is a singular point in Theorem~\ref{thmSecond-Order-Asymptotics}
and Theorem~\ref{thmFinite-Sample-Size}; however, we will see that
this represents a very trivial
case within classical hypothesis testing. Using Lemma~\ref{lemmaCQ-Duality}, we check that the
equivalent conditions of $V(\rho\|\sigma)=0$ is as follows: (i) $\rho
$ and $\sigma$ commute. This
means that $\rho$ and $\sigma$ can be simultaneously diagonalized as
$\rho=\sum_x\lambda(x)\vert a_x\rangle\langle a_x \vert$
and $\sigma=\sum_x\mu(x)\vert a_x\rangle\langle a_x \vert$,
and our problem reduces to a
classical one with probability
laws $\{\lambda(x)\}_x$ and $\{\mu(x)\}_x$. (ii) There is a constant
$k$ such that for all $x$ with
$\lambda(x)\neq0$, we have $\lambda(x)=k\mu(x)$; and actually $\log
k=D(\rho\|\sigma)$. Assigning
arbitrarily any $x^n$ with nonzero $\lambda^n(x^n)$ to the null
hypothesis $\rho^{\otimes n}$, we
obtain the best tradeoff between the type~I  error $\alpha_n$
and type~II error $\beta_n$,
and this is expressed as
$\alpha_n=1-\beta_n\exp\{nD(\rho\|\sigma)\}$.

\begin{appendix}\label{appA}
\section*{Appendix: Proof of lemmas}
We give proofs to the two lemmas presented in Section~\ref{subsectionproof-prep}.

\begin{pf*}{Proof of Lemma~\ref{lemmaCQ-Duality}}
This is done by direct calculation. For functions $v$~and~$w$, it is
obvious that
\[
\operatorname{Tr}v(\rho)=\sum_xv \bigl(\lambda(x)
\bigr)=\sum_{xy}v \bigl(\lambda(x) \bigr)|
\gamma_{xy}|^2
\]
and
\begin{eqnarray*}
\operatorname{Tr}v(\rho)w(\sigma) &=&\operatorname{Tr} \biggl(\sum
_xv \bigl(\lambda(x) \bigr)\vert a_x\rangle
\langle a_x \vert \biggr) \biggl(\sum
_yw \bigl(\mu(y) \bigr) \vert b_y
\rangle \langle b_y \vert \biggr)
\\
&=&\sum_{xy}v \bigl(\lambda(x) \bigr)w \bigl(\mu(y)
\bigr)|\gamma_{xy}|^2.
\end{eqnarray*}
Using these two equations with proper $v$ and $w$ at every step when
needed, we get
%
\begin{eqnarray}
\label{eqlemmCD-1} \operatorname{Tr}\rho(\log\rho-\log\sigma) &=&\sum
_{xy} \bigl(\lambda(x)\log\lambda(x)|\gamma_{xy}|^2-
\lambda(x)\log\mu(y)|\gamma_{xy}|^2 \bigr)
\nonumber\\[-8pt]\\[-8pt]
&=&\sum_{xy}P_{X,Y}(x,y)\log
\frac{\lambda(x)}{\mu(y)}=\expect \biggl(\log\frac{\lambda
(X)}{\mu(Y)} \biggr)\nonumber
\end{eqnarray}
and
%
\begin{eqnarray}
\label{eqlemmCD-2} && \operatorname{Tr}\rho(\log\rho-\log\sigma)^2\nonumber
\\
&&\qquad = \operatorname{Tr}\rho\log^2\rho-2\operatorname{Tr}(\rho\log\rho )\log
\sigma+\operatorname{Tr}\rho\log^2\sigma\nonumber
\\
&&\qquad =\sum_{xy} \bigl(\lambda(x)\log^2
\lambda(x)|\gamma_{xy}|^2
\\[-4pt]
&&\hspace*{48pt}{} -2\lambda(x)\log\lambda(x)
\log\mu(y)|\gamma_{xy}|^2+\lambda(x)\log^2\mu(y)|\gamma_{xy}|^2 \bigr)\nonumber
\\
&&\qquad =\sum_{xy}P_{X,Y}(x,y) \biggl(\log
\frac{\lambda(x)}{\mu(y)} \biggr)^2 =\expect \biggl(\log\frac
{\lambda(X)}{\mu(Y)}
\biggr)^2.\nonumber
\end{eqnarray}
Equation~(\ref{eqlemmCD-1}) confirms equation~(\ref{eqR-E-E}), and
equations~(\ref{eqlemmCD-1})
and (\ref{eqlemmCD-2}) together lead to equation~(\ref{eqR-V-V}).
Thus we finish the proof
of Lemma~\ref{lemmaCQ-Duality}.
\end{pf*}

\begin{pf*}{Proof of Lemma~\ref{lemmavec-proj-near}}
We show equation~(\ref{eqlemma-vpn-1}) as follows:
\begin{eqnarray*}
\hspace*{-5pt}&& \bigl\| \bigl(\vert\phi\rangle \langle\phi\vert \bigr)|\varphi \rangle
\bigr\|^2- \bigl\| \bigl(\pi\vert\phi\rangle \langle\phi\vert\pi \bigr)|\varphi
\rangle\bigr\|^2
\\
\hspace*{-5pt}&&\qquad =  \bigl( \bigl\| \bigl(\vert\phi\rangle \langle\phi\vert \bigr)|\varphi \rangle\bigr\|+
\bigl\| \bigl(\pi\vert\phi\rangle \langle\phi \vert\pi \bigr)|\varphi \rangle\bigr\| \bigr)
\times \bigl( \bigl\| \bigl(\vert \phi\rangle \langle\phi\vert \bigr)|\varphi \rangle
\bigr\|- \bigl\| \bigl(\pi \vert\phi \rangle \langle\phi\vert\pi \bigr)|\varphi \rangle\bigr\|\bigr)
\\
\hspace*{-5pt}&&\qquad \leq 2 \bigl( \bigl\| \bigl(\vert\phi\rangle \langle\phi\vert \bigr)| \varphi
\rangle\bigr\|- \bigl\| \bigl(\pi\vert\phi\rangle \langle\phi\vert \pi \bigr)|\varphi
\rangle\bigr\| \bigr)
\\
\hspace*{-5pt}&&\qquad \leq 2 \bigl\| \bigl(\vert\phi\rangle \langle\phi\vert \bigr)|\varphi \rangle-
\bigl(\pi\vert\phi\rangle \langle\phi\vert\pi \bigr)|\varphi \rangle\bigr\|
\\
\hspace*{-5pt}&&\qquad =  2 \bigl\| \bigl( \langle\phi|(\mathbh{1}-\pi)|\varphi \rangle \bigr)
\pi| \phi\rangle+ \bigl(\langle\phi| \varphi\rangle \bigr) (\mathbh {1}-\pi)|\phi
\rangle \bigr\|
\\
\hspace*{-5pt}&&\qquad = 2\sqrt{ \bigl| \langle\phi|(\mathbh{1}-\pi)|\varphi \rangle
\bigr|^2\cdot\bigl\|\pi|\phi\rangle\bigr\|^2+ \bigl|\langle\phi| \varphi
\rangle\bigl|^2\cdot\bigl\|(\mathbh{1}-\pi)|\phi\rangle\bigr\|^2}
\\
\hspace*{-5pt}&&\qquad \leq 2\sqrt{ \bigl\||\phi\rangle-\pi|\phi\rangle\bigr\|^2\cdot1 + 1\cdot \bigl\||
\phi\rangle-\pi|\phi\rangle\bigr\|^2}
\\
\hspace*{-5pt}&&\qquad \leq 2\sqrt{\varepsilon^2+\varepsilon^2}=2\sqrt{2}
\varepsilon,
\end{eqnarray*}
where the fourth line is by the triangle inequality, the sixth line is
due to Pythagoras'
theorem and the other lines are trivially by direct calculations and
the conditions stated
in the lemma.
\end{pf*}
\end{appendix}

\section*{Acknowledgments}
The author would like to thank Fernando Brand\~{a}o, Shunlong Luo,
William Matthews, Mil\'an Mosonyi and Andreas Winter for interesting
discussions. He is
especially grateful to Masahito Hayashi for many helpful discussions.



\printaddresses

\end{document}